\documentclass{SolarPhysics}    
\usepackage[optionalrh]{spr-sola-addons}
\usepackage{graphicx}	
\usepackage{epstopdf}
\usepackage{epsfig}
\usepackage{url}
\usepackage{fixltx2e}
\usepackage[usenames]{color}
\normalsize

\newcommand{\ang}{\AA\ }
\newcommand{\gapprox}{\lower.4ex\hbox{$\;\buildrel >\over{\scriptstyle\sim}\;$}}
\newcommand{\lapprox}{\lower.4ex\hbox{$\;\buildrel <\over{\scriptstyle\sim}\;$}}
\newcommand{\arcsec}{\hbox{$^{\prime\prime}$}}

\def\ang{\AA}
\def\etal{{\it et al.,~}}

\begin{document}
\begin{article}
\begin{opening}
\title{The Compatibility of Flare Temperatures Observed with AIA, GOES, and RHESSI}
\runningtitle{Solar Flare Temperatures}

\author{Daniel F.\ Ryan$^{1,2,3}$, Aidan M. O'Flannagain$^1$, Markus J. Aschwanden$^4$, and Peter T.\ Gallagher$^1$}
\runningauthor{Ryan \etal}

\institute{$^1$School of Physics, Trinity College Dublin, Dublin 2, Ireland\\
	   $^2$Solar Physics Laboratory (Code 671), Heliophysics Science Division, NASA Goddard Space Flight Center, Greenbelt, MD 20771, USA\\
	   $^3$Catholic University of America, Washington, DC 20064, U.S.A.	   
	   $^4$Solar and Astrophysics Laboratory,
	   Lockheed Martin Advanced Technology Center, 
           Dept. ADBS, Bldg.252, 3251 Hanover St., Palo Alto, CA 94304, USA; 
           (e-mail: \url{aschwanden@lmsal.com})}

\date{Received ... ; Revised ... ; Accepted ...}

\begin{abstract}
We test the compatibility and biases of multi-thermal flare DEM (differential emission measure) peak temperatures determined with AIA with those determined by GOES and RHESSI using the isothermal assumption.  In a set of 149 M- and X-class flares observed during the first two years of the SDO mission, AIA finds DEM peak temperatures at the time of the peak GOES 1--8~\ang\ flux to have an average of $T_{\rm p} = 12.0 \pm 2.9 $ MK and Gaussian DEM widths of $\log_{10}(\sigma_{\rm T}) = 0.50 \pm 0.13$.  From GOES observations of the same 149 events, a mean temperature of $T_{\rm p} = 15.6 \pm 2.4 $ MK is inferred, which is systematically higher by a factor of $T_{\rm GOES}/T_{\rm AIA}=1.4\pm0.4$. We demonstrate that this discrepancy results from the isothermal assumption in the inversion of the GOES filter ratio.  From isothermal fits to photon spectra at energies of $\epsilon \approx$~6--12~keV of 61 of these events, RHESSI finds the temperature to be higher still by a factor of $T_{\rm RHESSI}/T_{\rm AIA}=1.9\pm1.0$.  We find that this is partly a consequence of the isothermal assumption. However, RHESSI is not sensitive to the low-temperature range of the DEM peak, and thus RHESSI samples only the high-temperature tail of the DEM function.  This can also contribute to the discrepancy between AIA and RHESSI temperatures.  The higher flare temperatures found by GOES and RHESSI imply correspondingly lower emission measures.  We conclude that self-consistent flare DEM temperatures and emission measures require simultaneous fitting of EUV (AIA) and soft X-ray (GOES and RHESSI) fluxes.  
\end{abstract}


\end{opening}

\section{Introduction}
\label{sec:intro}
The temperature of the solar corona is one of its most fundamental characteristics.  It affects the nature of its physical processes and properties such as radiation, conduction, waves, shocks, the plasma-$\beta$, hydrodynamics, {\it etc.}  One of the most notable phenomena which encompasses many of these processes is solar flares.  Flares are believed to occur when energy stored in stressed magnetic fields is suddenly released, causing, among other things, a rapid heating of the flare plasma.  Temperature measurements play a vital role in better understanding these eruptive events.  Observational studies of flare thermal energy budgets ({\it e.g.}, \opencite{emsl12}), thermodynamic properties ({\it e.g.}, \opencite{feld96b}; \opencite{ryan12}), hydrodynamic scaling laws ({\it e.g.}, \opencite{rtv78}; \opencite{asch08}), flare cooling ({\it e.g.}, \opencite{raft09}; \opencite{ryan13a}) as well as many others all depend on temperature measurements.  In order to perform these measurements, an array of satellite instruments has been developed.  Among these are the {\it X-Ray Sensor} onboard the {\sl Geostationary Operational Environmental Satellites} (GOES/XRS), the {\sl Reuven Ramaty High Energy Solar Spectroscopic Imager} (RHESSI, \opencite{lin02}), and the {\it Atmospheric Imaging Assembly} onboard the {\sl Solar Dynamics Observatory} (SDO/AIA, \opencite{leme12}).  These instruments are sensitive to most of the temperature range in which coronal flare plasmas are typically found, 0.5--20~MK (AIA), $\approx$4--40~MK (GOES) and $\approx$7--100~MK (RHESSI).

However, in order to understand both the context and limitations of temperature measurements made with these instruments, it is important to know how they compare and the cause of any discrepancies.  Previous studies have compared the temperature measurements of GOES and RHESSI and typically found that RHESSI exhibits systematically higher temperatures.  \inlinecite{batt05} computed RHESSI peak temperatures of 85 B-class -- M-class flares from an isothermal fit (or two isothermal fits) convolved with a non-thermal fit.  The temperature of the isothermal fit (or cooler isothermal fit), $T_1$, was compared to the GOES temperature, $T_{\rm G}$.  The GOES temperature was found to be systematically lower.  A loose relationship was found and fitted by $T_1 = 1.12T_{\rm G}+3.12$.  This implies that for flares of temperatures of 10--25~MK, typical of M- and X-class flares \cite{ryan12}, RHESSI gives higher temperatures than GOES by 4--6~MK.  \inlinecite{mcti09} compared RHESSI and GOES temperature measurements of the non-flaring Sun from 2002--2006.  He found that the average RHESSI temperature was 6--8~MK while the average GOES temperature was 4--6~MK.  This is broadly consistent with \inlinecite{batt05}.  These measurements of \inlinecite{mcti09} result in a temperature ratio of $T_{\rm RHESSI} / T_{\rm GOES}$ = 1.4$\pm$0.2.  In the same year, \inlinecite{raft09} examined the temperature evolution of a C1.0 flare with several instruments including GOES and RHESSI.  The maximum RHESSI temperature was found to be $\approx$15~MK, while the maximum GOES temperature was found to be 10~MK.  This results is a temperature ratio of $T_{\rm RHESSI} / T_{\rm GOES}$~=~1.5 which is slightly higher than previous studies.  However, the RHESSI maximum temperature was found to occur $\approx4$~minutes before the GOES maximum.  Taking these temperature measurements simultaneously would lower this ratio, bringing it more into line with previous studies.

In this paper we calculate the GOES and RHESSI temperatures of an ensemble of M- and X-class flares using the an isothermal assumption (as in previous studies).  We compare them to the peak temperature of the differential emission measure distribution (DEM) calculated with AIA, as per \inlinecite{asch13a} and \inlinecite{asch13b}.  In doing so, we explore the effect of the traditional isothermal assumption on the GOES and RHESSI temperature measurements and quantify the resulting bias.  In Section~\ref{sec:data}, we discuss the instrumentation, observations and data analysis of AIA, GOES and RHESSI.  In Section~\ref{sec:disc} we devise theoretical predictions of the effect of the isothermal assumption on GOES and RHESSI temperatures as compared to DEMs of various widths.  These predictions are then compared to the discrepancies between the peak temperatures of AIA DEM peak and those from GOES and RHESSI.  Finally in Section~\ref{sec:concl} we provide our conclusions.

\section{Data Analysis}
\label{sec:data}
\subsection{SDO/AIA Measurements}
\label{sec:aia}
AIA takes full disk images of the Sun in ten UV and EUV wavelength passbands, six of which are sensitive to coronal flaring plasma (335, 211, 193, 171, 131, and 94~\ang).  AIA operates with a cadence of $\Delta t=12$ s for a full set of images in all wavelengths.  The pixel size of the images is $\Delta x = 0.6\arcsec$ and the spatial resolution is $2.5 \Delta x = 1.5\arcsec \approx 1100$ km.  During the first two years of the SDO mission 155 M- and X-class flares were observed.  These were analyzed in a single-wavelength study at 335 \ang\ by \inlinecite{asch12}.  Multi-wavelength studies using all six coronal filters have analyzed the spatio-temporal parameters of these flares \cite{asch13b}, as well as their temperature and DEMs \cite{asch13c}. In this study we utilize the AIA DEM analysis method of \inlinecite{asch13b} which represents the peak of the DEM distribution as a single Gaussian.  This can be characterized by three parameters: the DEM peak emission measure $EM_{\rm p}$, a DEM peak temperature $T_{\rm p}$, and a Gaussian width $\log_{10}{(\sigma_{\rm T})}$ (see Equation~(\ref{eqn:dem}) in Section~\ref{sec:goesbias}).  This method was initially applied to the AIA observations of all the 155 M- and X-class flares at the time of the peak in the GOES 1--8~\AA\ flux.  Successful DEM fits were not found for five of these flares (flare numbers 11, 18, 35, 90, and 100 in \opencite{asch13c}) while sufficient GOES observations were not available for one additional flare (flare number 69 in \opencite{asch13c}).  Thus 149 of the 155 M- and X-class flares were analysed as part of this study.

\begin{figure}
\centerline{\includegraphics[width=1.0\textwidth]{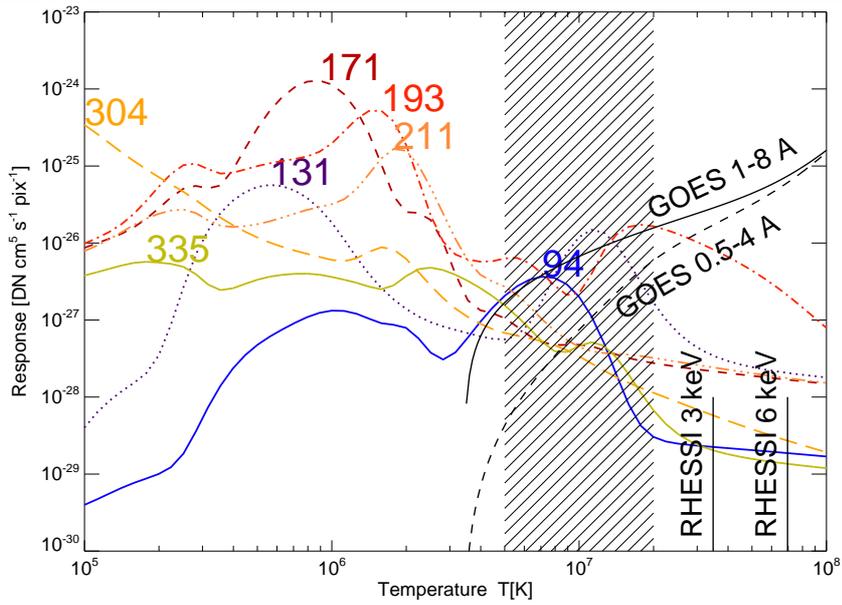}}
\caption{Temperature-response functions for the six coronal EUV channels along with the 304\,\ang\, channel of the {\sl Atmospheric Imaging Assembly (AIA)} onboard the {\sl Solar Dynamics Observatory}, according to the status of Dec 2012.  The GOES 1-8 \ang\ and 0.5-4 \ang\ is also shown (in arbitrary flux units), as well as thermal energy of the lowest fittable RHESSI channels at 3 keV and 6 keV. The approximate peak temperature range of large flares ($T_{\rm p} \approx 5-20$ MK) is indicated with a hatched area.}
\label{fig:tempresp}
\end{figure}

\begin{figure}
\centerline{\includegraphics[width=1.0\textwidth]{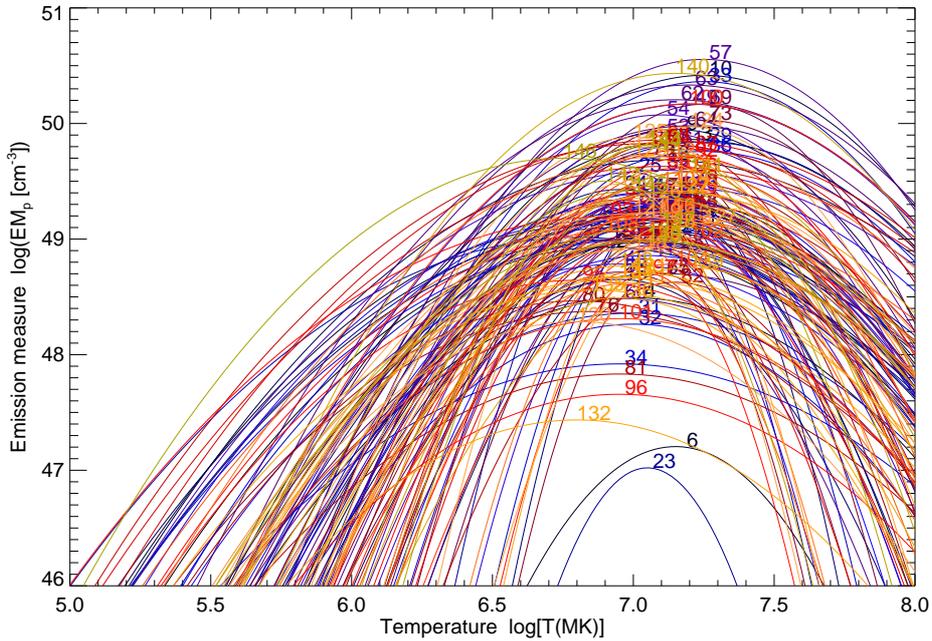}}
\caption{Gaussian DEM fits of the 149 M- and X-class flares analyzed in this study.}
\label{fig:dems}
\end{figure}

A key point in comparing temperatures measurements from different instruments is the simultaneity of observations. Since we used 1-min averaged GOES time profiles, the simultaneity between AIA and GOES is $(t_{\rm GOES}-t_{\rm AIA}) \approx 0.5 \pm 0.5$ min. Another decisive criterion is the temperature coverage. From the AIA response functions shown in Figure~\ref{fig:tempresp} we can see that the AIA filters have their primary or secondary peaks in the range from $\log_{10}(T) \gapprox 5.8$ (131 \ang ) to $\log_{10}(T) \lapprox 7.3$ (94, 131, and 193 \ang ), which is the temperature range in which a DEM distribution can be reliably obtained.  A display of all 149 single-Gaussian fits to the AIA data is shown in Figure~\ref{fig:dems}. The peak emission measures, integrated over the total flare volume, are found in the range of $\log_{10}(EM_{\rm p})=47.0-50.5$, with a mean and standard deviation of $\log_{10}(EM_{\rm p}) = 49.2 \pm 0.6$, in units of cm$^{-3}$. The flare peak temperatures are found in the range of $T_{\rm p}=5.6-17.8$ MK, with a mean and standard deviation of $T_{\rm p}=12.0 \pm 2.9$ MK.  The Gaussian half widths are found in the range of $\log_{10}(\sigma_{\rm T})=0.50 \pm 0.13$, which corresponds to a temperature factor of $10^{0.5} \approx 3.2$. Since a single-Gaussian function has only three parameters, the DEM fitting to six coronal filters is a very robust procedure and we are confident that the peak emission measure and peak temperature are, for the most part, accurately retrieved.  However, in a few cases it may not yield an acceptable $\chi^2$-value of the fit (see Table 2 in \opencite{asch13c}).  The next best option would be a four-parameter function.  Such a function could comprise of two semi-Gaussians joined together at the DEM peak with two different widths, $\sigma_{T1}$ at the low-temperature side, and $\sigma_{T2}$ at the high-temperature side ({\it e.g.}, as used in \opencite{asch01}).  While AIA may not provide sufficient temperature coverage to constrain the high-temperature side at $T \gapprox 20$ MK, RHESSI could provide strong constraints in this high-temperature tail. On the other hand, RHESSI does not have sufficient temperature coverage to constrain the peak temperature on the low-temperature side of the DEM, as we will see in Section 3.2.

\begin{figure}
\centerline{\includegraphics[width=1.0\textwidth]{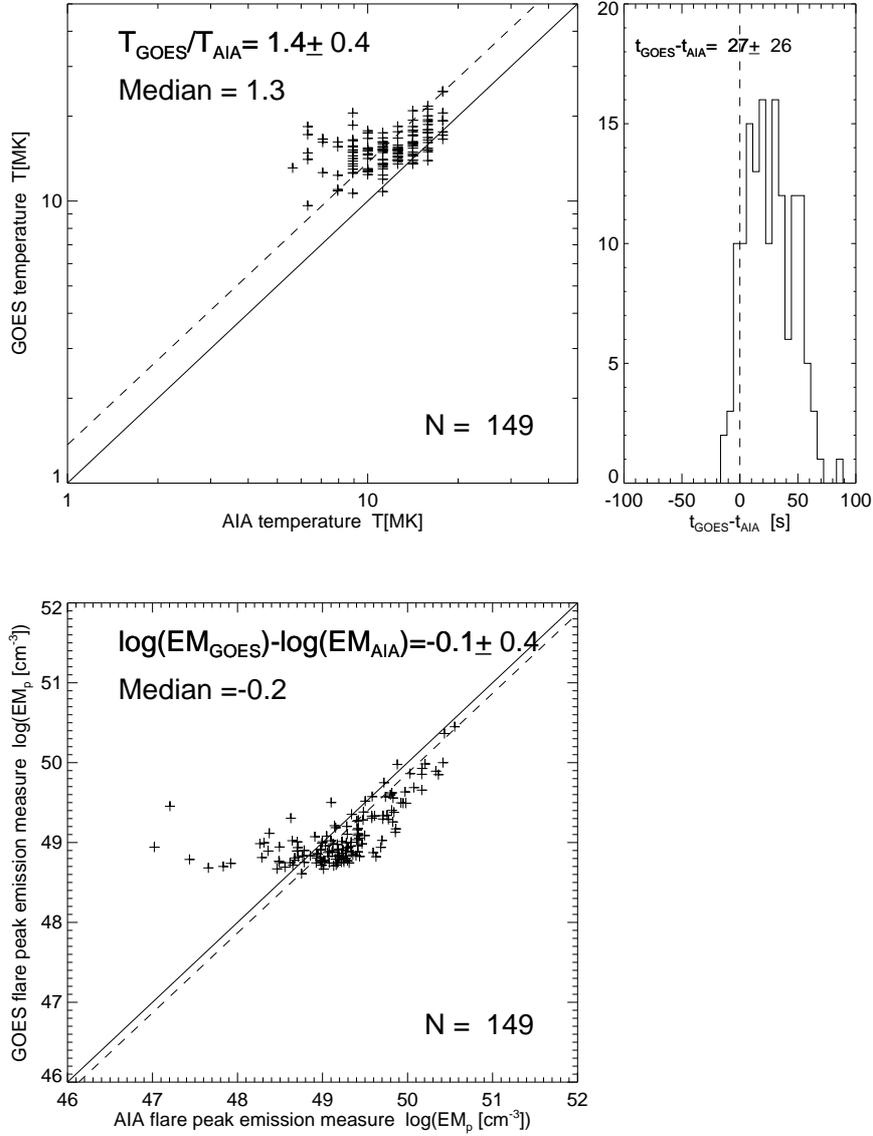}}
\caption{GOES versus AIA peak temperatures (top left panel) and peak emission measures (bottom left panel). The flare peak times refer to the GOES long wavelength (1--8 \ang ) peak time, $t_{\rm GOES}$, and coincides with the times of AIA measurements, $t_{\rm AIA}$, within the used time resolution of $\approx 1$ min.  See the histogram of time differences in top right panel, which has a mean and standard deviation of $(t_{\rm GOES}-t_{\rm AIA})=27 \pm 26$ s.}
\label{fig:goesvsaia}
\end{figure}

\subsection{GOES Measurements}
\label{sec:goes}
The GOES observations in this study were made by the XRSs onboard the GOES-14 and -15 satellites.  The XRS observes spatially integrated solar X-ray flux in two wavelength bands (long; 1--8~\AA, and short; 0.5--4~\AA) every two seconds.  By assuming the emitting plasma is isothermal, its temperature and emission measure can be calculated from the ratio of these channels \cite{whit05}.  In doing so, coronal abundances \cite{feld92}, the ionization equilibria \cite{mazz98}, and a constant density of 10$^{10}$~cm$^{-3}$ are assumed.  This final assumption was justified by \inlinecite{whit05}.  They used CHIANTI to calculate the spectra of isothermal plasmas at 10~MK with densities of $10^9$, $10^{10}$, and $10^{11}$ cm$^{-3}$ and found no significant differences.  The GOES channels have temperature sensitivities in the range $\approx$4--40~MK, as seen in Figure~\ref{fig:tempresp}.  It is therefore blind to the cooler coronal plasma which dominates the response functions of several of the AIA filters.  However, it is well suited to observing the peak temperatures of M- and  X-class flares which GOES typically finds to be between 10--25~MK \cite{ryan12}.

To ensure the accuracy of the GOES temperatures, a background subtraction must be performed to remove the influence of non-flaring plasma.  This was done using the Temperature and Emission measure-Based Background Subtraction method (TEBBS; \opencite{ryan12}).  This automatically finds the most suitable background subtraction based on how characteristic the resultant temperature and emission measure behavior is of a solar flare.  It then uses the method of \inlinecite{whit05} to find the flare temperatures and emission measures.

The GOES temperatures and emission measures resulting from the TEBBS analysis were found for the same 149 M- and X-class flares observed by AIA.  The top right panel of Figure~\ref{fig:goesvsaia} is a histogram of the time difference between the GOES long channel peak, $t_{\rm GOES}$, and the AIA measurements, $t_{\rm AIA}$.  This reveals that the average difference is 27$\pm$26~s.  Thus the condition for the simultaneity of measurements is satisfied.

The top left panel of Figure~\ref{fig:goesvsaia} shows the GOES temperature of each event plotted against the peak DEM temperature found with AIA.  A positive correlation is evident.  Comparison with 1:1 line (solid) reveals that the GOES temperatures are systematically higher than those found with AIA.  The GOES temperatures range from $\approx$10--24~MK with a mean and standard deviation of 15.6$\pm$2.4~MK.  This corresponds to an average ratio and standard deviation of $T_{\rm GOES}/T_{\rm AIA}$ of 1.4 $\pm$ 0.4 (dashed line).  This agrees visually with the distribution which, despite three or four flares with particularly low AIA temperatures relative to GOES, shows that the vast majority of points lie within a factor of two of the average trend.

The bottom panel of Figure~\ref{fig:goesvsaia} shows the GOES emission measure as a function of the DEM peak emission measure as calculated with AIA.  This distribution also shows a positive correlation with well confined scatter.  Once again there are three or four events which show a deviation from the trend at low AIA emission measures.  However these do not significantly affect the distribution.  The GOES emission measures range from 10$^{48.6}$--10$^{50.5}$~cm$^{-3}$ with a mean and standard deviation of 10$^{49.1\pm0.4}$~cm$^{-3}$.  Once again the mean is represented by the dashed line.  Comparing this to the 1:1 line (solid), it can be seen that GOES emission measures are systematically lower than the AIA values and imply an average ratio of $EM_{GOES} / EM_{AIA}$ = 10$^{-0.1 \pm 0.4}$, or $\approx$0.8.




\subsection{RHESSI Measurements}
\label{sec:rhessi}
RHESSI is capable of producing solar X-ray spectra in the range 3~keV to 17~MeV, with a spectral resolution of $\approx$1~keV in the range 3--100~keV \cite{smi02}. X-rays of energies 3--$\sim$25~keV are generally thermal bremsstrahlung originating from solar plasma at temperatures of $\approx$7~MK and above. By making the assumption that this plasma is isothermal, a temperature and emission measure can be produced by fitting a model thermal spectrum to RHESSI observations. Coronal plasma temperatures indicated by RHESSI studies vary from 5--10~MK during quiet-sun conditions \cite{mcti09} to 10--15~MK during microflares \cite{han08}. However, flares exhibiting a super-hot plasma component have been studied, with derived temperatures reaching up to 40~MK \cite{lin81,cas10}.

\begin{figure}
\centerline{\includegraphics[width=1.0\textwidth]{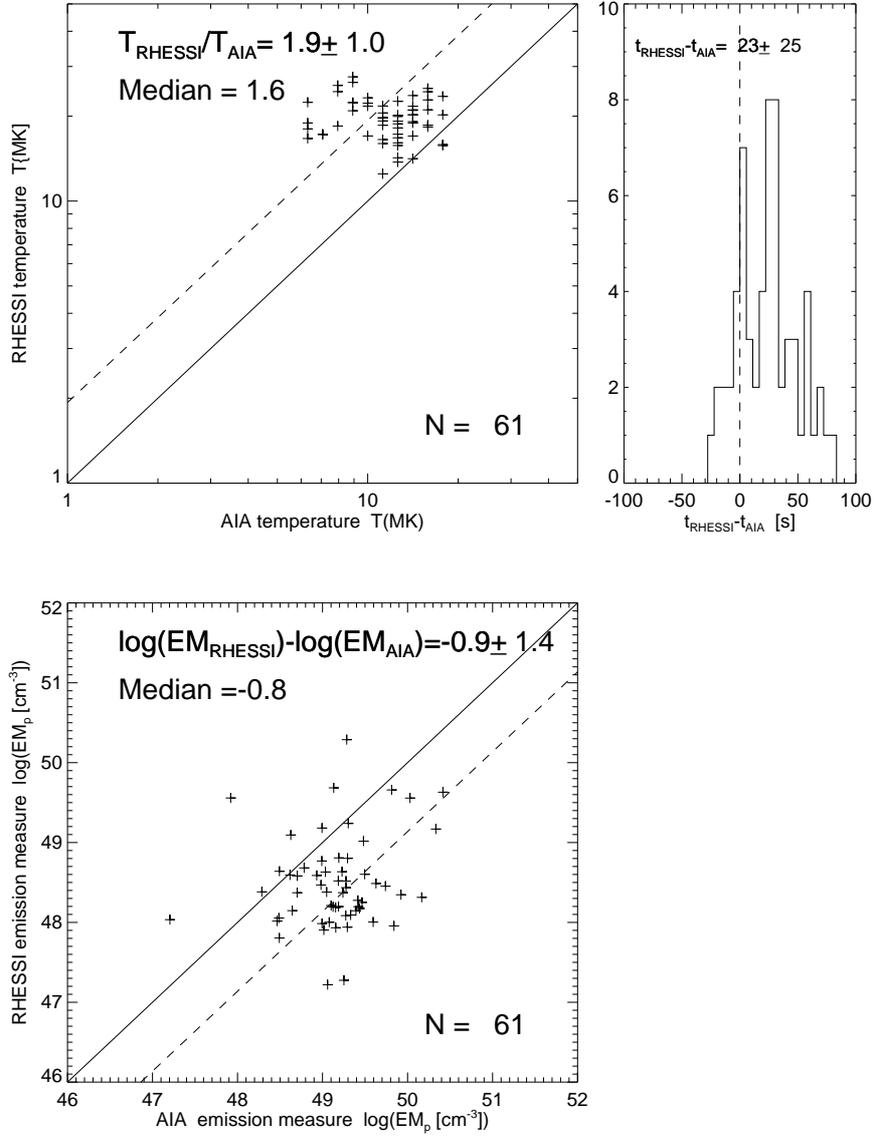}}
\caption{RHESSI versus AIA peak temperatures (top left panel)
and peak emission measures (bottom left panel). The flare peak times 
refer to the GOES long wavelength (1--8 \ang ) peak time, $t_{\rm GOES}$,
and coincides with the times of AIA measurements, $t_{\rm AIA}$, within
the used time resolution of $\approx 1$ min.  See the histogram of
time differences in top right panel, which has a mean and standard
deviation of $(t_{\rm RHESSI}-t_{\rm AIA})=23 \pm 25$ s.}
\label{fig:rhessivsaia}
\end{figure}

Of the 149 M- and X-class events used in this study, 61 were well observed by RHESSI. The remaining events occurred during RHESSI's passage through the South Atlantic Anomaly or the shadow of the Earth. A further number of events occurred during the annealing process of RHESSI's detectors, carried out in January and February of 2012. For the remaining events, we performed systematic fitting procedures to the spectra for the eight second time interval surrounding the time of peak GOES emission. In order to ensure that only the thermal component was included in the fitting process, the energy range of the fit was set to 5--20~keV for all intervals. As all of the studied flares were M- or X-class, RHESSI's aluminium attenuators were automatically moved in front of the grids to protect the germanium detectors during periods of peak flux. This meant that the only valid spectra to be used for background subtraction were those taken during adjacent night intervals, when solar emission was occulted by the Earth. However, as the count rate during these events was so high above quiet-sun background, the requirement for subtraction was vastly diminished.  The top right panel of Figure~\ref{fig:rhessivsaia} shows a histogram of the difference between measurement times of the RHESSI and AIA observations.  Once again it peaks below one minute within uncertainty demonstrating that the requirement for measurement simultaneity is satisfied.


From the 61 flare spectra analyzed, the temperatures were found to have a mean and standard deviation of $T_{\rm RHESSI} =21\pm10$~MK. This value is higher than both the GOES-derived (15.6$\pm$2.4~MK), and AIA DEM peak values (12.0$\pm$2.9~MK). The average ratio between the RHESSI and GOES temperatures was found to be $T_{\rm RHESSI} / T_{\rm GOES}$ = 1.3$\pm$0.7, which agrees very well with \inlinecite{batt05}, \inlinecite{mcti09}, and \inlinecite{raft09}.  The RHESSI temperatures are plotted against the AIA DEM peak temperatures in the top panel of Figure~\ref{fig:rhessivsaia}.  Comparison of the average temperature ratio (dashed line) with the 1:1 line (solid line) confirms that RHESSI exhibits higher temperatures with an average temperature ratio of $T_{\rm RHESSI} / T_{\rm AIA}$ = 1.9$\pm$1.0.  However, in contrast to the GOES distribution, no clear increasing trend is visible in this distribution.  This suggests that the RHESSI temperature is not closely related to the DEM peak measured with AIA, but rather skewed towards the high-temperature tail of the DEM.  

A similar scenario is seen in the bottom panel of Figure~\ref{fig:rhessivsaia} which shows RHESSI emission measures as a function of AIA peak DEM values.  Here the mean RHESSI emission measure was found to be roughly $10^{49.0}$~cm$^{-3}$.  This is systematically lower than both the GOES and AIA values which have averages of $10^{49.1}$ and $10^{49.2}$~cm$^{-3}$ respectively, and corresponds to an emission measure ratio of $EM_{RHESSI} / EM_{AIA}$ = 10$^{-0.9}$ = 0.13.


\section{Discussion}
\label{sec:disc}
\subsection{The GOES Temperature Bias}
\label{sec:goesbias}
In order to understand the discrepancies between the DEM peak temperatures obtained with AIA and GOES, we have to investigate the effect of multi-thermal DEMs on the GOES filter ratio.  The standard GOES temperature and emission measure inversions \cite{thom85,whit05} are based on the assumption of an isothermal plasma, which corresponds to a $\delta$-like DEM.  AIA has 6 coronal channels that constrain the DEM, and we assume here that a Gaussian DEM distribution (in $\log_{10}$T-space) fitted to these fluxes yields an acceptable approximation of the peak emission measure and temperature of the true DEM.

\begin{figure}
\centerline{\includegraphics[width=1.0\textwidth]{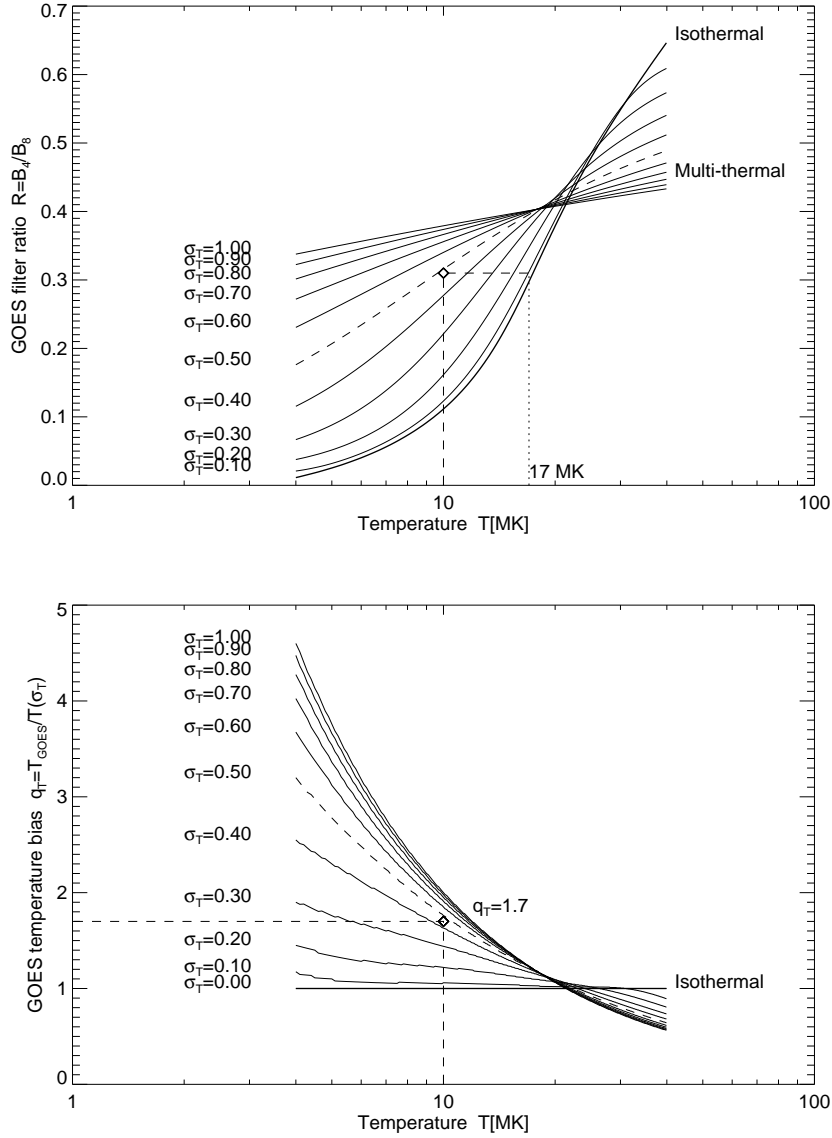}}
\caption{Top: The filter ratio of the GOES 0.5--4 \ang\ 
to the 1--8 \ang\ channel is shown for an isothermal DEM (thick curve)
and for Gaussian DEM distributions with Gaussian widths of
$\log_{10}(\sigma_{\rm T})=0.1,...,1.0$. The filter ratio is $R = B_4/B_8=0.31$ for
an isothermal DEM with a peak at $T_{\rm p}=10$ MK. For a Gaussian
DEM with a width of $\sigma_{\rm T}=0.5$ (dashed curve), the corresponding
isothermal filter-ratio corresponds to a temperature of $T_{\rm p}=17$~MK,
which defines a temperature bias of $q_{\rm T}=T_{iso}/T_{\sigma_{\rm T}}=1.7$.
Bottom: The temperature bias of multi-thermal DEMs with a
peak temperature at $T_{\rm p}(\sigma_{\rm T})$ compared with the temperature,
$T_{\rm iso}$, of isothermal DEMs is shown as a function of the
temperature and for a set of Gaussian widths, $\sigma_{\rm T}$.}
\label{fig:goesbias}
\end{figure}

For the GOES response functions, we use the simple expressions from the original fits of \inlinecite{thom85}. Updated and more complicated expressions specified with separate sets of polynomial coefficients for each of the GOES spacecraft are given in \inlinecite{whit05}.  However, these are expected to yield very similar results. The temperature-dependent part of the GOES long channel response function, $b_8(T)$, can be fitted with a third-order polynomial with temperature, $T$, in units of MK (Equation~(10) in \opencite{thom85}),
\begin{equation}\label{eqn:goesfiltfit}
	10^{55} b_8(T) =
	-3.86 + 1.17 T - 1.31 \times 10^{-2} T^2 + 1.78 \times 10^{-4} T^3 .
\end{equation}
The temperature itself can be expressed as a function of the ratio of the GOES short ($B_4$) and long ($B_8$) channel fluxes, $R(T)=B_4(T)/B_8(T)$.  This is equivalent to the ratio of the temperature dependent parts of the response functions, $R(T)=b_4(T)/b_8(T)$.  The relation between temperature and the GOES filter ratio is then given by (Equation~(9) in \opencite{thom85}),
\begin{equation}\label{eqn:goest}
	T(R) = 3.15 + 77.2 R - 164 R^2 + 205 R^3.
\end{equation}
Using Equations~(\ref{eqn:goesfiltfit}) and (\ref{eqn:goest}), the emission measure, $EM$, can then be derived from the measured long channel flux.
\begin{equation}\label{eqn:goesem}
	EM = B_8 / b_8(T) .
\end{equation}
The GOES filter ratio as a function of the temperature can easily be inverted from Equation~(\ref{eqn:goest}) by numerical interpolation of {\it R}-values for a fixed temperature array, $T(R_i)$, in the range of $0 < R < 1$. This GOES filter ratio $R(T)$ is shown in Figure~\ref{fig:goesbias} (curve labeled `isothermal' in top panel) and varies from $R(T=4$ MK) $\approx 0.01$ to $R(T=40$ MK) $\approx 0.66$. 

We can calculate the GOES filter ratio for Gaussian DEM distributions (in $\log_{10}$T) with particular values for the Gaussian width, $\sigma_{\rm T}$.  This is done by convolving the Gaussian DEM distributions,
\begin{equation}\label{eqn:dem}
 	{dEM(T) \over dT} = EM_{\rm p} \ \exp{\left( 
		{-[\log_{10}(T)-\log_{10}(T_{\rm p})]^2 \over 2 \sigma_{\rm T}^2} \right)} \ .
\end{equation}
The GOES short and long channel fluxes can be then directly computed with the GOES response functions $\rho_8(T)=b_8 10^{55}$ and $\rho_4(T) = \rho_8(T) \times R(T)$,
\begin{equation}\label{eqn:goesshort}
	B_4 = \int{ {dEM(T) \over dT} \rho_4(T) dT } \ ,
\end{equation}
\begin{equation}\label{eqn:goeslong}
	B_8 = \int{ {dEM(T) \over dT} \rho_8(T) dT } \ .
\end{equation}
From this, the GOES filter ratios, $R(T,\sigma_{\rm T})$, for any arbitrary temperature width, $\sigma_{\rm T}$, can be trivially obtained. These multi-thermal GOES filter ratios are shown in Figure~\ref{fig:goesbias} (top panel) for a range of widths, $\sigma_{\rm T}=0.1,...,1.0$.  The slope of the filter ratio progressively flattens for larger thermal widths $\sigma_{\rm T}$. For instance, the GOES filter ratio $R(T=10$ MK, $\sigma_{\rm T}=0)\approx 0.11$ for an isothermal DEM at a temperature of $T_{\rm p}=10$ MK, but increases to $R(T=10$ MK, $\sigma_{\rm T}=0.5) \approx 0.31$ for a Gaussian width of $\sigma_{\rm T}=0.5$ (marked with a dashed line in Figure~\ref{fig:goesbias}). Consequently, if we make the assumption of an isothermal plasma, as it is done in the standard application of GOES-derived temperatures, we would infer from the same observed filter-ratio $R=0.31$ an isothermal temperature of $T_{\rm GOES}~=~17$~MK.  We would thus overestimate the peak DEM temperature by a factor of $q_{\rm T}=T_{\rm GOES}/T_{\rm p}(\sigma_{\rm T}=0.5)=1.7$. 

These temperature bias factors, $q_{\rm T}=T_{\rm GOES}/T_{\rm p}(\sigma_{\rm T})$, are computed for a number of Gaussian widths in the range of $\log_{10}(\sigma_{\rm T})=0.1,...,1.0$ in Figure~\ref{fig:goesbias} (bottom panel).  From these calculations we see that the GOES temperatures are generally overestimated for flare peak temperatures of $T_{\rm p} \lapprox 22$ MK, while they are underestimated above this critical value.  The critical value $T_{\rm crit}\approx 22$ MK is related to an inversion point in the GOES isothermal filter ratio function $R(T)$.  The overestimation can be as large as a factor of four for low flare temperatures near $T \gapprox 4.0$ MK and for broad multi-thermal DEMs with a Gaussian width of $\sigma_{\rm T} \approx 1.0$.

This temperature bias, $q_{\rm T}$, can approximately be fitted by
\begin{equation}\label{eqn:goesbiasfit}
	q_{\rm T} = {T_{\rm GOES} \over T(\sigma_{\rm T})}
	\approx \left({ 22 \over T_{MK}}\right)^{0.9 \sigma_{\rm T}} \ .
\end{equation}
This relation was determined empirically.  Nonetheless it was found to satisfactorily reproduce the relationship between temperature bias, DEM temperature peak, and DEM width as calculated more rigorously using Equations~(\ref{eqn:goesfiltfit})\,--\,(\ref{eqn:goesshort}) ({\it i.e.}, bottom panel of Figure~\ref{fig:goesbias}).
For the particular data set of 149 M- and X-class flares observed with AIA in this study, we measured a mean DEM peak temperature of $T_{\rm AIA} = 12.0 \pm 2.9$ MK and Gaussian DEM half widths of $\log_{10}(\sigma_{\rm T})=0.50 \pm 0.14$.  From this we predict (with Equation~(\ref{eqn:goesbiasfit})) a mean GOES temperature bias of 
\begin{equation}\label{eqn:goesbiaspred}
	q_{\rm T}^{\rm pred} = 1.4 \pm 0.3 \ .
\end{equation}
When rounded to one decimal place, this precisely matches the observed GOES to AIA temperature ratio
\begin{equation}\label{eqn:goesbiasobs}
	q_{\rm T}^{\rm obs} = 1.4 \pm 0.4 \ .
\end{equation}
The residuals between observed and predicted temperature ratios ($q_{\rm T}^{\rm pred}-q_{\rm T}^{\rm obs}$) were found to be independent of temperature, suggesting that these averages well represent the overall distribution.

From these results, we conclude that GOES overestimates the peak temperature of large GOES flares (M- and X-class) on average by 40\%. Only for flare temperatures around $T_{\rm p} \approx 20$ MK the GOES temperature matches the DEM peak temperature.  For our sample we predict GOES temperatures with a mean of $T_{\rm GOES}^{\rm pred} = q_{\rm T} \times T_{\rm AIA} = 16.2 \pm 2.1$ MK, which also agrees with the observed temperature range of $T_{\rm GOES}=15.6 \pm 2.4$. 

\subsection{The RHESSI Temperature Bias}
\label{sec:rhessibias}
The temperature-dependent response functions (Figure~\ref{fig:tempresp}) show that the temperature range of AIA filters covers DEM peak temperatures of $T_{\rm AIA}\approx 0.5-20$~MK.  RHESSI covers $T_{\rm RHESSI} \approx 7-140$ MK, if we associate the fitted thermal energies of $\epsilon \approx 6-12$ keV with the DEM peak temperatures. This means that AIA and GOES can constrain the peak of flare DEMs well for flare temperatures of $T_{\rm p} \approx 4-20$ MK, while RHESSI applies thermal fits to the high-energy tail of the DEM distribution, but cannot constrain the peak of the DEM well. RHESSI fits to the thermal spectrum are often made with the assumption of an isothermal DEM.  However, the RHESSI data clearly show evidence that all flare DEMs cover a broad temperature range and therefore should be fitted with a multi-thermal DEM model ({\it e.g.}, \opencite{asch07}). In the following we will investigate the discrepancy in flare DEM peak temperatures resulting from isothermal RHESSI fits in the 6--12 keV range and multi-thermal (Gaussian) DEM fits obtained with AIA.

\begin{figure}
\centerline{\includegraphics[width=1.0\textwidth]{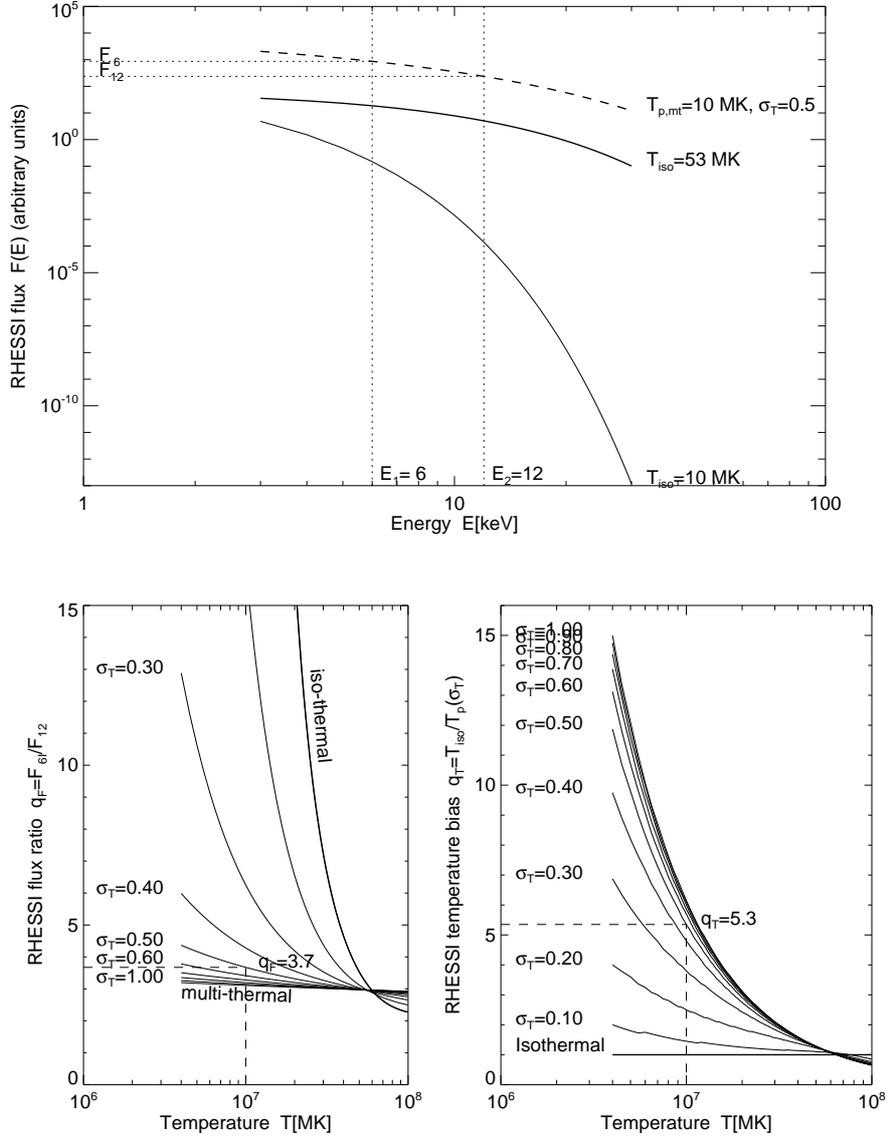}}
\caption{Top: Simulated RHESSI flux of a thermal photon spectrum
with an isothermal temperature of $T_{iso}=10$ MK (thin curve),
a multi-thermal spectrum with a peak temperature of $T_{MT}=10$
MK and a Gaussian width of $\log_{10}(\sigma_{\rm T})=0.5$ (dashed curve), 
and an isothermal spectrum that has the same flux ratio 
$q_F=F_6/F_{12}=3.7$, which is found for $T_{iso}=53$ MK, 
which corresponds to a temperature bias of $q_{\rm T}=T_{\rm RHESSI}/T_{\rm AIA}=5.3$.
Bottom left: The RHESSI flux ratio of isothermal and multithermal
spectra is shown as a function of the DEM peak temperature $T_{\rm p}$
for Gaussian DEM distributions with Gaussian widths of
$\log_{10}(\sigma_{\rm T})=0.1,...,1.0$. The flux ratio $q_F=3.7$ corresponding 
to the case shown in the top panel is marked with dashed line.
Bottom right: The temperature bias $q_{\rm T}=T_{iso}/T_{\rm p}$ 
of isothermal DEMs with a peak temperature at $T_{\rm p}$ is shown 
as a function of the peak temperature, $T_{\rm p}$, and for a set of 
Gaussian widths, $\sigma_{\rm T}$. The case with a temperature bias of
$q_{\rm T}=5.3$ of the spectrum shown in the top panel is indicated with
a dashed line.}
\label{fig:rhessibias}
\end{figure}

The bremsstrahlung spectrum $F(\epsilon)$ as a function of the photon energy $\epsilon~=~h\nu$ of an isothermal plasma with temperature, $T$, is (\opencite{brow74,dulk82}),
\begin{equation}\label{eqn:bremspec}
        F(\epsilon) = F_0 \int
        {\exp{(-{\epsilon / k_B T})} \over T^{1/2}} {dEM(T) \over dT} \ dT \ ,
\end{equation}
where $F_0 \approx 8.1 \times 10^{-39}$ keV s$^{-1}$ cm$^{-2}$ keV$^{-1}$.  This equation assumes the coronal electron density is equal to the ion density $(n=n_{\rm i}=n_{\rm e})$, the ion charge number $Z\approx 1$, and neglects factors of order unity, such as from the Gaunt $g(\nu, T)$.  The $dEM(T)/dT$ specifies the DEM ($n^2 dV$) in the element of volume $dV$ corresponding to temperature range $dT$,
\begin{equation}
        \left({dEM(T) \over dT}\right) dT = n^2(T) \ dV \ .
\end{equation}
Here we use the same parameterization of the DEM as in Section~\ref{sec:goesbias} (Equation~(\ref{eqn:dem})).  This can be characterized by three parameters, DEM peak emission measure $EM_{\rm p}$, DEM peak temperature $T_{\rm p}$, and Gaussian width $\log_{10}(\sigma_{\rm T})$, all of which we obtained in Section~\ref{sec:aia}.  Inserting the DEM function (Equation~(\ref{eqn:dem})) into the bremsstrahlung spectrum (Equation~(\ref{eqn:bremspec})) we obtain an isothermal spectrum for $\sigma_{\rm T} \mapsto 0$, and a multi-thermal spectrum for $\sigma_{\rm T} > 0$. As an example we show the isothermal ($T_{\rm p}=10$ MK) photon energy spectrum, $F(\varepsilon)$, in the energy range of $\varepsilon=3-30$ keV in Figure~\ref{fig:rhessibias} (top panel). We see that the thermal spectrum falls off steeply, with a flux ratio of $q_F=F_6/F_{12}=10^3$ between 6 keV and 12 keV.

Now we calculate a multi-thermal spectrum $F(\varepsilon)$ for a Gaussian DEM with the same peak temperature $T_{\rm p}=10$ MK, but a Gaussian width of $\log_{10}(\sigma_{\rm T})=0.5$.  This is shown in Figure~\ref{fig:rhessibias} (top panel, dashed spectrum).  It is much flatter and has a flux ratio of $q_F=F_6/F_{12} \approx 3.7$ between 6 keV and 12 keV.  This is more than two orders of magnitude smaller than the isothermal case.  An isothermal fit to this flux ratio would correspond to a DEM peak temperature of $T_{iso}=53$ MK because this temperature produces the same flux ratio of $q_F=F_6/F_{12} \approx 3.7$ (Figure~\ref{fig:rhessibias}, top panel, thick solid line). Thus the assumption of isothermal DEMs leads to significant overestimates of temperature and emission measure.  In the example here, the DEM peak temperature is overestimated by a factor of $q_{\rm T}=T_{iso}/T_{\rm p}(\sigma_{\rm T}=0.5)$=(53 MK/10 MK)=5.3, and the DEM peak emission measure is underestimated by about a factor of 0.03.  Since RHESSI spectra are often fitted with an isothermal spectrum, the obtained temperature virtually always overestimates the DEM peak temperature substantially.

Next we calculate the flux ratios, $q_F=F_6/F_{12}$, for a range of DEM Gaussian widths, $\log_{10}(\sigma_{\rm T})=0.1,...,1.0$, and show their dependence on the DEM peak temperature $T_{\rm p}$ (Figure~\ref{fig:rhessibias}, bottom left panel). The flux ratio is highest for an isothermal spectrum, but progressively decreases with broadening DEMs ({\it i.e.}, larger Gaussian widths, $\sigma_{\rm T}$). We also calculate the RHESSI isothermal temperature bias, $q_{\rm T}=T_{iso}/T_{\rm p}(\sigma_{\rm T})$, between an isothermal fit and a multithermal DEM (Figure.~\ref{fig:rhessibias}, bottom right). We see that the temperature overestimation can be up to a factor of $q_{\rm T}\approx 5$ for narrowband DEMs with $\log_{10}(\sigma_{\rm T})=0.25$ and low flare temperatures of $T_{\rm p} \approx 4$ MK, and up to the same factor for broadband DEMs with $\log_{10}(\sigma_{\rm T}) \approx 0.5-1.0$ for larger temperatures of $T_{\rm p} \approx 10-12$ MK, which are typically
measured in flares. 

Applying this model for the isothermal bias of spectral fits in the $\varepsilon=6-12$~keV energy range to the AIA flare measurements, we can predict the expected temperature range measured by RHESSI for the same set of M- and X-class flares. This isothermal temperature bias $q_{\rm T} = T_{\rm RHESSI}/T_{\rm p}(\sigma_{\rm T})$, shown in Figure~\ref{fig:rhessibias} (bottom right panel), can approximately be represented by the simple relationship,
\begin{equation}\label{eqn:rhessibiasfit}
	q_{\rm T} = {T_{\rm RHESSI} \over T_{\rm p}(\sigma_{\rm T})}
	\approx \left({ 60 \over T_{\rm p}(\sigma_{\rm T})}\right)^{(\sigma_{\rm T}^{1/2})} \ .
\end{equation}
This relation was determined empirically and, as in the case of Equation~(\ref{eqn:goesbiasfit}), was found to satisfactorily reproduce the relationship between temperature bias, DEM temperature peak and DEM width (bottom panel of Figure~\ref{fig:rhessibias}).
For the 61 flares observed by both AIA and RHESSI, we measured a mean DEM peak temperature of $T_{\rm AIA} = 12.0 \pm 2.9$ MK and Gaussian DEM half widths of $\log_{10}(\sigma_{\rm T})=0.51 \pm 0.14$.  Thus, using Equation~(\ref{eqn:rhessibiasfit}), we predict a mean RHESSI temperature of $T_{\rm RHESSI}=37.2 \pm 6.1$ MK, or a RHESSI isothermal temperature bias of $q_{\rm T}=T_{\rm RHESSI}/T_{\rm AIA}$ of 
\begin{equation}
	q_{\rm T}^{\rm pred} = 3.3 \pm 1.0 \ .
\end{equation}
This is commensurable with the observed RHESSI to AIA temperature ratio (Figure~\ref{fig:rhessivsaia}, top panel)
\begin{equation}
	q_{\rm T}^{\rm obs} = 1.9 \pm 1.0 \ .
\end{equation}
There is not a very close agreement between the observed and predicted temperature ratios.  However a very accurate prediction is not expected.  This is because the high-temperature part of the DEM in the range of $T_{\rm p} \approx$ 10--20~MK is not so well constrained with AIA, to which only the 193~\ang\ line (with a Fe {\sc xxiv} line) and the 94~\ang\  filters are sensitive. Also the shape of the DEM function, for which we choose a simple symmetric Gaussian, may not adequately describe the high-temperature tail of the DEM function.  This is supported by both the results of \inlinecite{grah13} and our finding that the residuals between observed and predicted temperature ratios ($q_{\rm T}^{\rm pred}-q_{\rm T}^{\rm obs}$) have a slight temperature dependence.  The residuals are greater for lower peak DEM temperatures.  For lower peak temperatures, the high-temperature part of the DEM sampled by RHESSI is further away (in temperature space) from the peak.  Therefore, the prediction of a temperature bias requires a greater extrapolation of the Gaussian DEM into the high-temperature tail.  This can exaggerate any discrepancy between the high-temperature tail predicted by a Gaussian parameterization and the `true' high-temperature tail sampled by RHESSI.  An asymmetric DEM function with a steeper fall-off at the high-temperature tail ({\it e.g.}, \opencite{asch01}) could bring the predicted RHESSI bias in better agreement with the observed RHESSI/AIA temperature ratio.  Despite this, our model prediction of a substantial temperature overestimation by isothermal fits to the RHESSI spectra is consistent with the systematically higher measured RHESSI temperatures.  Thus we conclude that self-consistent flare temperatures and emission measures require simultaneous fitting of EUV (AIA) and soft X-ray (GOES, RHESSI) fluxes with a suitably parameterized DEM distribution function.

\section{Conclusions}
\label{sec:concl}
In this paper, the differential emission measures (DEMs) of 149 M- and X-class flares were calculated at the time of the GOES peak 1--8~\ang\ flux using AIA.  GOES temperatures and emission measures of these events were also calculated at the flare peak using an isothermal assumption \cite{whit05,ryan12} and compared to the peak temperatures and emission measures of the AIA DEMs.  It was found that, on average, the GOES temperatures were a factor of 1.4$\pm$0.4 higher than the AIA DEM peak temperatures.  The temperatures and emission measures of 61 of these flares were also calculated with RHESSI using an isothermal fit to the observed spectra between 5--20~keV.  The RHESSI temperatures were found to be higher than both GOES and AIA.  On average the RHESSI temperatures were a factor of 1.9$\pm$1.0 higher than AIA and a factor of 1.3$\pm$0.7 higher than GOES.  The ratio of RHESSI to GOES temperatures was found to agree with previous studies.  Conversely, the GOES emission measures were typically lower than the AIA DEM peak emission measures, while the RHESSI emission measures were found to be lower still.

The effect of the isothermal assumption on the calculation of the GOES temperatures was investigated.  It was found that DEMs of greater widths (more multithermal) increasingly altered the relationship between DEM peak temperature and GOES filter ratio.  For temperatures less than 22~MK, the isothermal assumption was predicted to result in higher derived GOES temperatures than multithermal DEMs.  However, for temperatures greater than 22~MK, the isothermal assumption was predicted to lead to lower derived GOES temperatures than the multithermal DEMs.  The resulting bias between temperatures derived from the isothermal assumption, $T_{\rm GOES}$, and a DEM of width of $\sigma_{\rm T}$, was described by Equation~(\ref{eqn:goesbiasfit}).  This resulted in a mean predicted isothermal bias for the events observed by AIA and GOES of 1.4$\pm$0.3.  This agreed well with the observed GOES/AIA temperature ratio of 1.4$\pm$0.4.

A similar analysis was performed on derived RHESSI temperatures.  It was found that in the range 4--50~MK, the isothermal assumption was predicted to lead to higher derived temperatures than those obtained with multithermal DEMs.  The discrepancy was described by Equation~(\ref{eqn:rhessibiasfit}).  This resulted in a mean predicted RHESSI isothermal bias for the 61 events observed by both AIA and RHESSI of 3.3$\pm$1.0.  This is commensurate with the observed RHESSI/AIA temperature ratio of 1.9$\pm$1.0 but is not in close agreement.  However a close agreement is not necessarily expected since the high temperature tail of the DEM in the RHESSI temperature range is not well constrained by AIA and a symmetric Gaussian may not be best suited to describing this high temperature tail.  Therefore, in order to self-consistently obtain flare temperatures, EUV (AIA) and soft X-ray (GOES and RHESSI) fluxes must be simultaneously fit with a suitably parameterized DEM distribution function, {\it e.g.}, a bi-Gaussian.

\acknowledgements
The authors would like to thank the following for supporting this research: NASA (contract NNG04EA00C of the SDO/AIA instrument to LMSAL), the Fulbright Association, Catholic University of America, and the Irish Research Council.  Thanks must also go to Richard A. Schwartz for his helpful discussions.

\bibliographystyle{spr-mp-sola}
\bibliography{mscor3rdrd1}

\IfFileExists{\jobname.bbl}{} {\typeout{}
\typeout{****************************************************}
\typeout{****************************************************}
\typeout{** Please run "bibtex \jobname" to obtain} \typeout{**
the bibliography and then re-run LaTeX} \typeout{** twice to fix
the references !}
\typeout{****************************************************}
\typeout{****************************************************}
\typeout{}}

\end{article}

\end{document}